\documentclass[11pt]{article}

\usepackage{cite}
\usepackage{amssymb}
\usepackage{amsmath}
\usepackage{bm}

\begin{document}

\title{Some Personal Reflections on Quantum Non-locality and the Contributions of John Bell.}
\author{B. J. Hiley\footnote{E-mail address b.hiley@bbk.ac.uk.}.}
\date{TPRU, Birkbeck, University of London, Malet Street, \\London WC1E 7HX.\\ \vspace{0.4cm}(5 Sept 2014) }
\maketitle

\begin{abstract}
I present the background of the Bohm approach that led John Bell to a study  of quantum non-locality  from which his famous inequalities emerged. I recall the early experiments done at Birkbeck with an aim to explore the possibility of `spontaneous collapse', a way suggested by Schr\"{o}dinger to avoid the conclusion that quantum mechanics was grossly non-local.  I also review some of the work that John did which directly impinged on my own investigations into the foundations of quantum mechanics and report some new investigations towards a more fundamental theory.

\end{abstract}

\section{Introduction}

My first encounter with quantum non-locality was in discussions with David Bohm when I joined him as an assistant lecturer at Birkbeck in the sixties. Although my PhD thesis  had been in condensed matter physics under the supervision of Cyril Domb,  I had always been fascinated and puzzled by quantum phenomena and when the opportunity to study the subject with Bohm came up, I took it.  I would listen to the discussions between Bohm and Roger Penrose, who was at Birkbeck at the time, and it soon became clear that  the prescribed interpretation of quantum mechanics that I was taught as an undergraduate left too many questions unanswered.  

At that time there was a very strange atmosphere in the physics community. I was often informed that there were no problems with the interpretation of the quantum formalism.  If I did not follow the well prescribed rules of the quantum algorithm, I would be `wasting my time'.  There was no alternative, just do it!  But Louis de Broglie and David Bohm had shown there was another way. However their views were strongly opposed by many, and as far as I could judge their reasons did seem to be based on mathematics or on logic, but on a preconceived notion that the formalism was, in principle, the best we could do.    The experimental data was bizarre when looked upon from the standpoint of classical physics, but the uncertainty principle somehow prevented us from examining actual process in detail.  Reality was somehow veiled~\cite {bde03}.
Where we condemned to use a mere algorithm or was there an underlying ontology? 

In the late sixties I learnt that there was another physicist working at CERN, a mecca of orthodoxy no less, who shared our worries about quantum mechanics and that was John Bell.  I used to meet him at various conferences in Europe where we would exchange ideas.  As a reflection of the attitudes of the time, one day John remarked that he felt he had to keep a low profile at these meetings because his CERN colleagues might disapprove!

 Our discussions would be centred round Bohm's work~\cite{db52}, 
 published in 1952.  His proposals had met, not merely with scepticism but also, surprisingly, with open hostility.  Yet all he had  shown was that if we focus on the real part of the Schr\"{o}dinger equation, we find an equation that has the same form as the classical Hamilton-Jacobi equation but with the addition of an extra term.  This term was called the `quantum potential'  and with this term one could account for quantum behaviour without giving up the notion of a particle following a trajectory.  More importantly, this could also be done without giving up the uncertainty principle.
 
The story of the Bohm approach is by now well known either through our book, ``The Undivided Universe"~\cite{dbbh93} or  Peter Holland's excellent book~\cite{ph95}.
Or alternatively, through an offshoot of what is now called ``Bohmian Mechanics"~\cite{ddst09}. 
  ``The Undivided Universe" contains his last words on his own approach as he died just as we finished it, so I do not need to go into the details of the approach here.   I want to concentrate on one aspect of the approach which inspired John Bell to re-consider the foundation of quantum theory, namely, the appearance of non-locality.  As John remarked about Bohm's work -``he had seen the impossible done", but it was the appearance of the non-locality in the original paper~\cite{db52} that took his attention and he went on  to reflect deeply on this phenomenon before he finally produced his famous inequality~\cite{jb64}.

\section{Non-locality and Bohm.}

To see how non-locality arises in the Bohm approach, we need to look at the two-body entangled state, with wave function $\psi(x_1, x_2, t)$, where $x_1$ and $x_2$ are the coordinates of the two particles at a given time, $t$.  Under the polar decomposition  $\psi(x_1, x_2, t)= R(x_1, x_2, t)\exp[iS(x_1, x_2, t)/\hbar]$, $R$ and $S$ being real, the real part of the Schr\"{o}dinger equation becomes what we call the quantum Hamilton-Jacobi equation,
\begin{eqnarray}
\partial_t S+\frac{(\nabla_1S)^2}{2m_1}+\frac{(\nabla_{2} S)^2}{2m_2} + Q+V=0.	\label{eq:2body}
\end{eqnarray}
The similarity with the two-particle classical Hamilton-Jacobi equation is clear, except for an additional term,
\begin{eqnarray}
Q_\psi(x_1, x_2, t)=-\frac{\hbar^2}{2m_1}\frac{\nabla_1^2R(x_1,x_2,t)}{R(x_1,x_2,t)}-\frac{\hbar^2}{2m_2}\frac{\nabla_2^2R(x_1,x_2,t)}{R(x_1,x_2,t)}.	\label{eq:QPE}
\end{eqnarray}
This is the quantum potential and it is clearly non-local since it depends on the positions of both particles, $x_1$ and $x_2$, at the {\em same} time.  Notice that if the wave function is a product $\psi_1(x_1,t)\psi_2(x_2,t)$, the quantum potential becomes local as can be easily seen by substitution into the Schr\"{o}dinger equation.

Although this result was clearly stated in Bohm's original paper~\cite{db52}, the main discussion focussed on whether the ideas proposed by Bohm violated the von Neumann theorem~\cite{vn55} 
which claimed to show that  it is not possible to reproduce the results of standard quantum mechanics by adding so called `hidden variables'.  But Bohm had not added any new variables;  he had simply {\em interpreted} the coordinate $x(t)$ that appears in the wave function as being the actual position of the particle,  and {\em interpreted} $\nabla S$ as the momentum of the particle, so   no new variables were added.   

Clearly this interpretation assumes a particle has, simultaneously, a well-defined position {\em and} a well-defined momentum so  surely this must violate the uncertainty principle?  Actually no, because the uncertainty principle only claims that it is not possible to {\em measure} simultaneously the position and momentum.  It does not claim that it is  impossible for the particle to {\em possess simultaneously} a position and a momentum.  Note that if simultaneous measurement is ruled out, then it is impossible to experimentally determine whether the particle actually has or does not have simultaneous values of $x$ and $p$.  The standard approach merely {\em assumes} a particle cannot possess well-defined values of these variables simultaneously, whereas Bohm {\em assumes} the particle can have simultaneous $x$ and $p$, but we do not know what the values are.

 So how does the uncertainty principle enter?  Here we follow Wheeler~\cite{jw91}
and assume that measurement does not merely reveal those values present,  the measuring instrument becomes active in the measuring process, producing the appropriate eigenvalue of the operator that is being measured.  In this participatory act, the complementary variables are changed in an uncontrollable way.  Thus the statistical element of a quantum process remains.

There is much evidence that the Bohm approach, either in the form discussed by Bohm and Hiley~\cite{dbbh93} and Holland~\cite{ph95} 
or in the form of its offshoot, Bohmian mechanics,  as discussed by  D\"{u}rr, Goldstein and Zanghi~\cite{dgz96} 
and D\"{u}rr and Teufel~\cite{ddst09}, 
does not lead to any internal inconsistencies and, in fact, contains all the same experimental predictions as does the standard approach, so there are no experimental differences between the two approaches.   However because particles are assumed to have well-defined positions at all times, the approach is able to bring out very clearly the non-locality that appears with entangled states.  

It is this feature that John Bell noticed and asked ``Are all theories, which attribute properties to local entities while reproducing the results of quantum mechanics, non-local?"  Before answering that question, let us consider another question, ``Were there any clues of non-locality in the usual formulation of quantum theory?"

\section{Non-locality in the Standard Quantum Formalism?}
 
In this section I would like to highlight some key arguments that suggested in fact that non-locality was at the heart of quantum phenomena.   It is not merely a feature that arises only when the real part of the Schr\"{o}dinger equation is considered.  To bring this out,
we begin with a quotation from a review paper by Dirac~\cite{pd73} in 1973,
\begin{quote}
For an assembly of the particles we can set up field quantities which do change in a local way, but when we interpret them in terms of probabilities of particles, we get something which is non-local.........
\end{quote}
He goes on
\begin{quote}
 I think one ought to say that the problem of reconciling quantum theory and relativity is not solved.  The concepts which physicists are using at the present time are not adequate. 
 \end{quote}
 We can go back much earlier to find unease at the  appearance of some form of non-locality.  History treats the famous Einstein, Podolsky, Rosen paper~\cite{epr35} 
 as the beginning of the debate of non-locality, but as the title emphasises,  their paper concentrated on the {\em completeness problem}, that is whether we need additional parameters (hidden variables) to complete the quantum formalism and rid it of its statistical features. The question of non-locality was not directly confronted.  
 In reaching their conclusion, EPR added a caveat, 
 \begin{quote}
 Indeed, one would not arrive at our conclusion if one insisted that two or more physical quantities can be regarded as simultaneous elements of reality {\em only when they can be simultaneously measured or predicted}. On this point of view, since either one or the other,  but not both simultaneously, of the quantities $P$ and $Q$ can be predicted, they are not simultaneously real.  This makes the reality of $P$ and $Q$ depend upon the process of measurement carried out on the first system, which does not disturb the second system in any way.
  No reasonable definition of reality could be expected to permit this.
 \end{quote}
 Thus they were ruling out the possibility that measurement could be non-local and participatory.
 
 One should note that Schr\"{o}dinger had pointed out in a much earlier paper that there was a problem with the quantum description of a two-body system~\cite{es27}.  
 He noticed that, in the presence of a small coupling between two systems, the state of system \# 1 becomes entangled with the state of system \# 2, no matter how far apart they are spatially.  He emphasised this again, but more clearly in a later paper  when he writes:
 \begin{quote}
 If for a system which consists of two entirely separated systems the representative (or wave function) is known, then the current interpretation of quantum mechanics obliges us to admit not only that by suitable measurements, taken on one of the two parts only, the state (or representative or wave function) of the other part can be determined without interfering with it, but also that, in spite of this non-interference, the state arrived at depends quite decidedly on what measurements one chooses to take -- not only on the results they yield~\cite{es36} .
 \end{quote}
 This control  of the state of a distant system by an experimenter was to Schr\"{o}dinger: 
 \begin{quote}
 ... rather discomforting that the theory should allow a system to be steered or piloted into one or the other type of state at the experimenter's mercy in spite of his having no access to it. 
 \end{quote}
 Schr\"{o}dinger's papers did not seem to attract much attention at the time.  Rather it was Bohr who dominated the debate with Einstein.  The key element of Bohr's response to what Einstein called ``spooky action at a distance" appears in at least two places with identical wording, wording that I believe John Bell found, to say the least, unclear.  The exact wording in question is
 \begin{quote}  
 From our point of view we now see that the wording of the above mentioned criterion of physical reality proposed by Einstein, Podolsky and Rosen contains an ambiguity as regards the meaning of the expression {\em without in any way disturbing a system.}  Of course there is no question of a mechanical disturbance of the system under investigation during the last critical stage of the measuring procedure.  But even at this stage there is essentially the question of an {\em influence} [my emphasis] on the very conditions which define the possible types of predictions regarding the future behaviour of the system.\cite{nb35} [p. 700]
 \end{quote}
 The puzzle is clear, ``What does the word `influence' actually mean?" Notice it is not a mechanical disturbance, but an `influence'.  That is an extraordinary word to use in twentieth century science.   
Could it be that Bohr had already anticipated the appearance of a term like the quantum potential in the real part of the Schr\"{o}dinger equation?  

It is the quantum potential in equation (\ref{eq:2body}) that clearly brings out the non-locality.  Thus there is a new quality of energy emerging only at the quantum level and it is this energy that is responsible for the non-locality that arises in entangled states.

 Schr\"{o}dinger's equation is non-relativistic, so could it be that this non-local effect is an artifact of a non-relativistic theory?
The problem with that claim is that  we know entanglement is a necessary and indispensable feature of the standard quantum formalism. For example, we know that in helium, the ground state wave function of the two electrons is an entangled state and this is necessary to produce the correct energy levels.   Any relativistic corrections will merely produce small modifications, fine structure effects and not overrule the general result.  Perhaps, as Schr\"{o}dinger argued, when the particles are separated by a sufficient distance, the non-locality {\em spontaneously localises}.  Schr\"{o}dinger writes:
\begin{quote}
 It seems worth noticing that the paradox could be avoided by a very simple assumption, namely \dots the
{\em phase relations} between the complex coupling coefficients become entirely lost as a consequence of the process of separation. This would mean that not only the parts, but the whole system, would be in the situation of a mixture, not of a pure state. It would not preclude the possibility of determining the state of the first system by {\em suitable} measurements in the second one or vice versa. But it would utterly eliminate the experimenter's influence on the state of that system which he does not touch.
\end{quote}
This `spontaneous localisation' process would remove the unwelcome feature that `steers or pilots the distant system'.
Clearly the theory has these unpleasant features, but are such effects found in Nature?  Surely we must now turn to experiment.  

\section{Spontaneous localisation-New Physics?}

\subsection{Existence of Entangled state for Separated systems.}

It was in the mid-sixties as, unknown to us, John Bell was developing his ideas on non-locality that David Bohm, experimentalist David Butt and I discussed the possibility of looking for evidence of spontaneous localisation experimentally.  Rather than using position and momentum, we decided to use the polarisation properties of $\gamma$-rays, and confront directly the spin version of the EPR argument originally introduced by Bohm~\cite{db51}.

At this stage we were not aware that John was already in the process of publishing his inequalities paper in a new journal which, ironically, ceased publication after the first one or two volumes~\cite{jb64}.
This all happened in the days before the internet, when ideas filtered down very slowly. 

Be that as it may, our interest was in the experimental verification of the existence of entangled states over macroscopic distances.  Specifically, would the entanglement remain as the particle detectors were moved further apart or would there be some form of as yet unknown phase randomisation as suggested by Schr\"{o}dinger himself? 

At that time, in the mid-sixties, the best experimental evidence for entanglement existing over larger distances $[\sim 0.5m]$ was supplied by Langhoff~\cite{hl60}.
This experiment involved measurements of the correlation between the planes of polarization of the correlated gamma-rays produced by the annihilation of the $s$-state positronium produced by some radioactive source.  Langhoff used two different sources, $Na^{22}$ and $Cu^{64}$ in his experiment. The normalized ket for a pair of $\gamma$-photons immediately after annihilation is
\begin{eqnarray}
|\psi\rangle=\frac{1}{\sqrt 2}\left [ |x_1\rangle|y_2\rangle- |y_1\rangle|x_2\rangle\right]		\label{eq:entwf}
\end{eqnarray}
where $|x\rangle$ and $|y\rangle$ are the kets of the individual $\gamma$-photons plane-polarized in the $x$ and $y$ directions, respectively (see Kasday~\cite{lk71}).   Clearly equation (\ref{eq:entwf})  is an entangled state.  The photons propagate in opposite directions along the $z$-axis.

After travelling about 0.5m, the emerging photons undergo Compton scattering in silicon crystals.  The scattered photons are then detected by a pair of scintillation counters.  Coincidences  between the two counters are recorded when the azimuthal angles of the two counters are identical ($\phi=0$) and when they differ by $90^{\circ}$ $(\phi=\pi/2)$.  For the wave function (\ref{eq:entwf}), this ratio of coincidences scattered through the same scattering angle $\theta$ is given by 
\begin{eqnarray}
\rho=\frac{N_{\phi=\pi/2}}{N_{\phi=0}}=1+\frac{2\sin^4\theta}{\gamma^2-2\gamma\sin^2\theta}	\label{eq:anisot}
\end{eqnarray}
where
\begin{eqnarray*}
\gamma=2-\cos\theta+\frac{1}{2-\cos\theta}.
\end{eqnarray*}
This gives a theoretical value $\rho=2.60$ at $\theta=90^{\circ}$, with a maximum of $2.85$ at $\theta=82^{\circ}$.  These are the values before corrections are made for the angular resolution of the apparatus and before corrections for the geometrical size of the equipment is taken into account. For the Langhoff experiment, these corrections produced a theoretical maximum value for $\rho=2.48\pm0.2$.  The two sets of experimental results gave $\rho_{exp}=2.50\pm0.03$ for positrons produced by $Na^{22}$ and $\rho_{exp}=2.47\pm0.07$ for positrons produced by $Cu^{64}$.  Thus there is a good agreement between theory and experiment for a 0.5m separation.

\subsection{Is the Distance Between Scattering Centres Large Enough?}

Clearly non-locality exists over such a separation but is this distance large enough  to make sure a randomising process of the phase relation between the two photons has had time to take place?  Again is this distance far enough to notice any reduction in the `influence' that Bohr talks about?   To answer these questions we need to speculate about what factors could maintain these correlations over larger distances.

The first thing that comes to mind is the notion of a coherence length.  It is well known that interference effects depend on this feature so it could be that we require the photons be separated by distances greater than their coherence length before we see any evidence of a change in the pair wave function.  Thus we need to ensure that the scattering centres of the two annihilation  $\gamma$-rays are separated by a distance  that is greater than their combined coherence lengths.

The decay of $s$-state electron-positron pairs in  $Cu^{64}$ has been shown to occur with only a one-component lifetime and is found to be $(191 \pm 3)$ps~\cite{phpj71},
giving a coherence length of 0.057 m. From this we can assume that the width of the wave packet associated with each photon should be less than about 0.12 m. However this width is a somewhat ambiguous notion and, therefore, it is necessary to explore situations in which the source-scatterer distances are an order of magnitude greater than this to ensure that the overlap of the wave packets is indeed negligible. This suggests a distance of over 1m is required to make absolutely sure that the $\gamma$-rays are sufficiently separated so that their individual coherent lengths do not overlap.  As we have already indicated, in the Langhoff experiment, the separation between the scatters was only about 0.5m so there is room for doubt, even though this doubt is small.

There is also another feature that we must consider, namely to ensure the  scattering events are {\em space-like separated}.  If this condition is satisfied and the pure state predictions are confirmed, then we can rule out the possibility of any local hidden-variable theory that might involve some form of signalling at luminal or sub-luminal velocities, as one of the quanta is being measured. 

Ruling out this possibility will obviously depend upon the resolving time of the detection counters. In the Langhoff~\cite{hl60} experiment, the resolving time of the scintillation counters was 5 ns. This gives a photon path length of 1.5 m implying that the separation of the detectors must be at least this distance to ensure the detection events are space-like separated.  Unfortunately as the separation in this case is only 0.5m the detection events are not space-like separated.  This again means that we cannot rule out this type of mechanism, no matter how unlikely.

Although in the  Langhoff experiment the  events were not space-like separated, some later results  obtained by Faraci {\em et al}~\cite{dfdg74} 
 reported that a  decrease in the anisotropy was found for  larger separation distances (5.6 m). Such a distance would satisfy both criteria discussed above    suggesting that there was some form of spontaneous localisation taking place, so Wilson, Lowe and Butt [WLB]~\cite{awjldb67}
 decided to repeat these experiments, but this time, systematically increasing the source-scatterer distances.  
 
Unfortunately as the separation increases, the detection efficiency of the scatterers decreases dramatically, varying approximately as the square of the scatterer dimensions and the square of the solid angle subtended by the second detector at the scatterer for a fixed source-polarimeter separation.  Thus there is a need to start with physically larger scatterers.
 Unfortunately with large scatterers, the geometric corrections become more difficult, which makes it hard to calculate absolute values of the anisotropy so that the final results will be less meaningful. Thus the WLB experiment was designed, not to measure absolute values of the anisotropy factors, but to look for changes in the anisotropy as the separation distances were systematically increased to space-like separations and  beyond.
 
To estimate the order of magnitude of change expected, WLB used a suggestion of Furry ~\cite{wf36} who argued that the final state will not be given by (\ref{eq:entwf}) but by the mixture 
\begin{eqnarray}
\mbox{either}\quad                                                                                                                                                                                                                                                                                                                                                                                                                                                                                                                                                                                                                                                                                                                                                                                                                                                                                                                                                                                                                                                                                                                                                                                                                                                                                                                                                                                                                                                                                                                                                                                                                                                                                                                                                                         |x_1\rangle|y_2\rangle\quad\mbox{or}\quad   |y_1\rangle|x_2\rangle.  	\label{eq:simple}
\end{eqnarray}
A more general final state was suggested by Bohm and Aharonov (1957), namely,\cite{dbya57}
\begin{eqnarray}
\rho'=\int\int P(\theta, \phi)|1,+\theta,\phi\rangle|2,-\theta,\phi\rangle\langle1,+\theta,\phi|\langle 2,-\theta,\phi| d\theta d\phi.		\label{eq:mixed}
\end{eqnarray} 	
In both cases the anisotropy should decrease to about 1.0.  With geometric corrections in all these cases the maximum anisotropy factor is considerably less than 2. Hence, if there is any form of spontaneous localisation process taking place, we would expect to find a significant decrease in the anisotropy as the source-scatterer distances were increased. 

The experimental results show that there is no detectable change in the anisotropy as the  distances between the scatterers  are systematically increased to  4.9m.    Two asymmetric arrangements of the detectors were also investigated,  the left hand scatterer was placed  0.6m from the centre, while the right hand one was placed at 1.6m from the centre.  In the second measurement,  the respective separations were chosen to be 1.5m and 2.45m.
 In neither case was any significant change in the anisotropy detected.    
  
 The distances chosen by WLB ensured that the detector separation was well beyond the coherence length criteria and the detection events were space-like separated.  The resolving time of the detectors used by WLB was 1 ns, giving a light length of 0.3m as against a final separation of 4.9m.  In these sets of experiments they were unable to confirm the results of Faraci {\em et al}~\cite{dfdg74}. The details of the WLB experiment will be found in their paper and will not be discussed further here. 
A later experiment was performed by Paramananda and Butt~\cite{vpdb87}, with the separation distance between the two scatterers extended to 23m.  Again no change in the anisotropy was seen.  
Today experiments show that these correlations hold for distances of up to 144km~\cite{ruaa07}.

\section{Quantum Non-locality.}

These experimental results clearly show  the possibility that entangled states spontaneously localise to mixed product states as suggested by Schr\"{o}dinger~\cite{es35} and Furry~\cite{wf36} 
must be ruled out.  This leaves us with either Bohr's original explanation using his notion of `influence', or with the Bohm interpretation where the appearance of the non-local polarisation correlations was maintained by the quantum potential energy.  Of course there may be  other explanations, but whichever way we choose to go,  
quantum mechanics contains an element of non-locality totally foreign to classical physics. Naturally the appearance of non-locality in physical theories has been strongly resisted as it appears to deny the possibility of doing science at all if, in the extreme, everything is inseparable from everything else. Fortunately not all states are entangled and there are many product states which are local.

From the above discussion, we see that the clearest indication of non-locality appears in the Bohm treatment of the quantum formalism.  In this approach, it is the re-introduction of the notion of a localised particle to which independent properties can be attached that shows clearly the presence of non-locality.  But attaching properties to particles immediately re-ignites the debate about `hidden variables' and the completeness problem raised by EPR.  In addition to this, criticisms have been raised on the grounds that it was a blatant attempt to return to the ideas of classical physics.  However this a curious criticism because it is this approach that highlights  non-locality, a feature that is totally foreign to classical thinking.

Fortunately John Bell~\cite{jb87} 
very quickly understood the significance of what Bohm had done~\cite{db52} writing,
\begin{quote}
Bohm's 1952 papers on quantum mechanics were for me a revelation\dots I have always felt since that people who have not grasped the ideas of those papers (and unfortunately they remain the majority) are handicapped in any discussion of the meaning of quantum mechanics.
\end{quote}
Bohm's approach, in Bell's own words, had replaced  ``unprofessionally vague and ambiguous" features of the standard approach with a ``sharpness that brings into focus some awkward questions".  In fact it was the appearance of non-locality in equation (\ref{eq:2body}) that triggered the question,  ``Will all theories that attach properties to individual particles and reproduce the results of quantum mechanics, be non-local?"

In a famous paper in 1964 his inequalities first appeared. These inequalities became  the focus of an intense theoretical debate as to their precise meaning and implication.  This discussion also produced a wealth of experiments showing that the inequalities were, in fact, violated, confirming the early results - the existence of non-locality in quantum phenomena.  A discussion of the inequality and its experimental consequences will be found elsewhere in this volume so I will not make any further comment here. 
 
 \section{Global properties and quantum non-locality}
 
 \subsection{Orthogonal groups, spin groups and the Clifford algebra}

 For me, John Bell's most important contribution to quantum physics has been his constant criticism of the standard interpretation of the formalism itself.  This criticism often echoed some of my own worries about the interpretation.  Contrary to classical physics, which concerns the description of what is actually going on in an evolving  physical process, quantum physics focusses on the measurement and what the observer `knows'.  Why should Nature care about what we, products of Nature, know or do not know in order to evolve?  
 
Traditionally physics tried to find descriptions that are entirely independent of the observer and, more importantly, what the observer knows or does not know about the unfolding process.  In quantum mechanics, we seem to have given up this tradition.  The standard interpretation talks about the results of measurement $A$, followed by the results of measurement $B$, but then declares that it is not possible talk about what goes on in between measurements.  In other words one only speaks about the {\em results of external intervention.}  Indeed as Bohr~\cite{nb61} writes,
\begin{quote}
As regards the specification of the conditions for any well-defined application of formalism, it is moreover essential that the {\em whole experimental arrangement} be taken into account.  In fact, the introduction of any further piece of apparatus like a mirror, in the way of a particle might imply new interference effects essentially influencing the predictions as regard the results to be eventually recorded.
\end{quote}
Bell came out very strongly against giving measurement such a prominent position~\cite{jb90}.
For him, the `measurement problem', which still plagues quantum theory, arises because the measuring instrument is singled out to play a special role. But surely any instrument is simply a collection of atoms governed by the very same laws that  we are investigating, so what makes it special?   All other physical processes end up in a linear superposition of eigenstates; the exception is the measuring instrument.  Why?
One essentially has to resort to a tautology:    `a measuring instrument is a system that reveals an eigenvalue of some operator'.  We seem to be arguing that what the measuring instrument does is to replace the logical `and' by the logical `or', without giving a physical reason to justify such a step.

Bell drew attention to two possible ways of avoiding this measurement problem.  The first way was to use the de Broglie-Bohm theory, the theory that leads to equation (\ref{eq:2body}).  Here the theory assumes the particles have sharp position values, $X(t)$, and these trace out a trajectory which the particle is assumed to follow.  Thus we have a way of characterising individual quantum processes.  The ensemble is characterised by a set of trajectories that are fully determined by the quantum  Hamilton-Jacobi equation, namely, the real part of the Schr\"{o}dinger equation.  The trajectory characterising a particular individual particle  is contingent on its initial position, a value that cannot be precisely controlled by the observer.

 The second way is to maintain that we don't need to worry about whether a particle follows a trajectory or not and to continue using the wave function which has an additional feature, namely, from time to time, it undergoes some new random collapse process as proposed by Ghirardi, Rimini and Weber\cite{grw86}.
In other words this collapse process can be regarded as a spontaneous spatial localisation of the micro-constituents occurring at random times, a process of which Schr\"{o}dinger would approve. The mean frequency of the localisation is assumed to be extremely small, with the localisation width being large on an atomic scale. In this way, no prediction of standard quantum formalism is changed in any appreciable way.

Clearly my own preference is the theory presented in Bohm and Hiley~\cite{dbbh93} simply because the experiments of Wilson, Lowe and Butt~\cite{awjldb67}, its extension by Paramananda and Butt~\cite{vpdb87}, together with a considerable number of other later experimental results, provide no evidence for any kind of spontaneous localisation over very large distances.  However there is one caveat: both theories are non-relativistic and it would be very enlightening to find some relativistic generalisation so that we can examine more closely how this non-locality sits in `peaceful coexistence' with relativity.

\subsection{Relativistic Considerations}

The obvious candidate for an investigation of a relativistic generalisation of the Bohm approach would be to separate the Klein-Gordon equation into its real and imaginary parts. However there are some serious difficulties  with this procedure which were discussed in detail in chapter 11 of Bohm and Hiley~\cite{dbbh93}.  To avoid these difficulties it was found necessary to go to the field theory before a satisfactory treatment of bosons in general could be found~\cite{dbbhpk87, pk94}. 
We will not go into this approach as it will take us into an area not directly relevant to the rest of this paper, so we will, instead, turn to consider fermions even though these also  present difficulties. 

For ferimions two new features appear, spin and relativity.  The preliminary attempt to apply the original method used by Bohm~\cite{db52} for the Schr\"{o}dinger equation to the Pauli equation~\cite{bst55} met with some success although the method did not inspire full confidence. Nevertheless some extremely illuminating illustrations of the model were published by 
Dewdney, Holland, Kyprianidis and Vigier~\cite{cdphak87, cdphak88}. 
In spite of these successes, the extension to the relativistic Dirac equation presented considerable difficulties although several early unsuccessful attempts were made~\cite{db62, ldb60, ph95, dbbh93}.  It was possible to replace the guidance condition\footnote{I prefer to call $p_\psi$ the Bohm momentum for reasons that will become clear in section 6.}, $p_\psi=\nabla S$ by the Dirac current (see, for example, Bohm and Hiley~\cite{dbbh93}) but no satisfactory relativistic generalisation of the quantum Hamilton-Jacobi equation (\ref{eq:2body}) was found even for the case of a single particle.

John Bell~\cite{jb87} himself attempted to rectify this situation by developing a quantum field theoretic approach, 
but his approach was not very satisfactory and his ideas have not been developed any further as far as I am aware. More recently, following on from the earlier work of Hestenes~\cite{dh03}
 I have, together with Bob Callaghan, developed an  approach to the Dirac equation using the full structure of the orthogonal Clifford algebra~\cite{bhbc12, bhbc11}.  We also applied the method to the Pauli equation removing some of the unsatisfactory features of the earlier attempts.  Fortunately we found that this attempt is a special case  so the results of Dewdney, Holland and Kyprianidis~\cite{cdphak87} still stand giving us the best illustration of non-locality for two non-relativistic particles with spin in an entangled state.
In this more general approach,  Hilbert space representations are not used, rather we make full use of the orthogonal Clifford algebra.  This means replacing the wave function by what can  essentially be regarded as a density matrix, $\rho$, but now defined algebraically in terms of elements of suitable left and right ideals. These elements contain all the information carried by the wave function and so describe the state of the system.  To anyone who had read the third edition of Dirac's classic book, {\em The Principles of Quantum Mechanics}~\cite{pd47} 
carefully, this will come as no surprise and explains the meaning of his {\em standard ket}.  Furthermore as Fr\"{o}hlich~\cite{hf67} has already pointed out, this density matrix is not the statistical matrix introduced by von Neumann~\cite{vn55}.
The formalism can even be used for a single particle in a pure state, in which   case $\rho^2=\rho$.  

John Bell~\cite{js80} himself attempted to use the density matrix in a different context, namely, to analyse  
the delayed-choice experiment and concluded that the density matrix could not be used in the de Broglie-Bohm theory because the theory gives fundamental significance to the wave function.  However it turns out that this conclusion is not correct.  The density matrix can be used but this means generalising the whole approach~\cite{bh13}.
I believe that it is the {\em insistence} on using the wave function that has held up progress in the development not only of a relativistic generalisation of the de Broglie-Bohm approach, but also standard quantum mechanics.  In the algebraic approach the wave function appears as a special case of a more general structure and is not basic.

In the algebraic approach, the choice of what ideals to use  is determined by the physics of the system under consideration.  This has been explained in detail in Hiley and Callaghan~\cite{bhbc11}.  Essentially the necessity of using the Clifford algebra for  the Pauli and Dirac equations arises because the wave function does not capture the full implications of the non-commutative structure  which only become significant in these algebras. To bring out these features we must go to the algebra.  In such algebras one must distinguish between left and right operations. This means that when we come to consider time development we need two equations, a left translation and a right translation-the equation and its dual.  In the case of the Schr\"{o}dinger equation, the dual is simply the complex conjugate equation, which is dismissed since it seems to  add nothing new.  This cannot be done for the Pauli or Dirac equations where the full non-commutative structure is necessary.

Given these two equations, if we subtract the equation from its dual, we  get immediately the quantum Hamilton-Jacobi equation.  On the other hand if we add the two equations, we get the Liouville equation.  By using this generalisation, we give a complete description of the one-particle dynamics of both the Pauli and the Dirac equations~\cite{bh13}.  We find in the generalised quantum Hamilton-Jacobi expressions for the quantum potential energy in all cases.
 Thus not only have we generalised the approach to a non-relativistic particle with spin, removing the unsatisfactory features of the Bohm-Schiller-Tiomno approach~\cite{bst55}, but we are able to obtain its relativistic generalisation, namely, the Dirac equation.   This immediately opens the way to treat fermion fields, a question in which Bell himself was interested.  

If we examine the expression of the Dirac quantum potential energy, we find it is a Lorentz scalar, a result I found rather surprising, although it provides a clue as to why quantum non-locality and relativity can exist in ``peaceful coexistence".  A scalar is frame independent and does not propagate. Furthermore in the Dirac case, we find  {\em two} currents appearing.  One is the Dirac current used in Bohm and Hiley~\cite{dbbh93}.  The other is the generalisation of the Schr\"{o}dinger current which gives rise to the guidance condition $p=\nabla S$.  This is a feature that has already been pointed out by Tucker~\cite{rt88} using a different approach to the Clifford algebra so that it is a general feature of the relativistic case with spin.
 The full implications of this unexpected result have still to be fully understood. 
 
 So far our study has been confined to the one-body problem.  Of much greater interest is the two-body problem, where the question of quantum non-locality will directly confront relativistic locality. Unfortunately the generalisation  is proving difficult at this present time.  This situation is very tantalising as the  the two Pauli particle case has been beautifully analysed by Dewdney {\em et al}~\cite{cdphak88}
However even in this non-relativistic case, I have not been able to extend the algebraic approach to the two-body system, but the investigation into the whole method is still on going at the time of writing.

In one sense the success achieved by using the orthogonal Clifford algebra should not be too surprising because the conventional wave functions of the Pauli and Dirac particles are simply Hilbert space representations of elements contained in the Clifford algebra itself.

One of the important features of the Clifford algebra is that it automatically contains the spin group, the double cover of the orthogonal group.  Normally this is treated by the somewhat abstract method of constructing spin bundles.  However the Clifford algebra was originally created, before the advent of quantum theory, to understand the properties of space and to discuss movement in space.  The Clifford group, the spin group, describes the global features of the rotation group, and gives a natural account of the difference between the $2\pi$ and $4\pi$ rotations.  This is a feature of the rotational properties of space and has, {\em a priori}, little to do with quantum phenomena.  It is rather physical processes exploit these global properties with quantum non-locality being merely a consequence of these global features.  This would account for the appearance of non-locality in spin correlations but what about the non-locality revealed in the $x-p$ correlations originally proposed by the EPR discussion?

\subsection{The Symplectic group, its double cover and another Clifford algebra}

If we are to understand the $x - p$ non-locality as a global feature in the same spirit as the rotational non-locality discussed in the previous section, we need to express the translation dynamics in an analogous algebraic structure. Thus while the orthogonal Clifford algebra is the geometric algebra of rotations, there ought to be a geometric algebra dealing with translations. 

At first sight this seems a nonstarter because translations in space form an Abelian group, but we want a non-commutative structure if we are to find an analogue of the orthogonal Clifford algebra.  One such candidate was first proposed by von Neumann way back in 1931~\cite{vn31}.  This is the famous paper that provides the foundation of the Stone-von Neumann theorem, the theorem that establishes the uniqueness of the Schr\"{o}dinger representation.  But we are not interested here in representations, we are interested in the algebraic structure itself.

Translations in space involve movement, that is momentum, so that not only must we consider translation in space, but we must consider  changes of momentum, that is translations in momentum space.  These sets of translations do not commute and it is this feature that gives rise to   a non-commutative symplectic space. This leads us into the rich field of symplectic geometry, a geometry that contains subtle topological structure having a very relevant significance for quantum processes.  I am deeply grateful to the mathematician Maurice de Gosson~\cite{mdg01, mdg10}, an expert in symplectic geometry, who has helped me understand some of the difficult aspects of the geometry.

For the purposes of this paper, I will give a brief account of this structure because John himself tried to use this approach to explore in more detail quantum non-locality.  Let us  start with von Neumann who, following Weyl~\cite{hw28}, 
 introduces a pair of non-commuting translation operators, $U(\alpha)=e^{i\alpha {\widehat P}}$ (translations in space) and $V(\beta)=e^{i\beta{\widehat X}}$ (translations in momentum space).  These operators satisfy the relations   
\begin{eqnarray}
U(\alpha)V(\beta) = e^{i\alpha\beta}V(\beta)U(\alpha),	\label{eq:Weyl}
\end{eqnarray}
together with
\begin{eqnarray}
U(\alpha)U(\beta)=U(\alpha + \beta);  \hspace{0.5cm}  V(\alpha)V(\beta)=V(\alpha + \beta). \nonumber
\end{eqnarray}						
von Neumann then defines an operator
\begin{eqnarray}
\widehat S(\alpha,\beta)=e^{i(\alpha\widehat P+\beta \widehat X)}=e^{-i\alpha\beta/2}U(\alpha)V(\beta)=e^{i\alpha\beta/2}V(\beta)U(\alpha)  \nonumber
\end{eqnarray}
and proves that the operator $\widehat S(\alpha, \beta)$ can  be used to define uniquely any bounded operator $\hat A$ on a Hilbert space  through the relation
\begin{eqnarray}
\hat A=\int\int a(\alpha, \beta)\widehat S(\alpha,\beta)d\alpha d\beta.
\label{eq:symA}				
\end{eqnarray}
Here $a(\alpha,\beta)$ is a function on a Schwartz space spanned by two variables $\alpha$ and $\beta$ in $\mathbb R^{2N}$. Thus we can establish a one-to-one relationship between the operator algebra of quantum mechanics and a set of real valued functions, $a(\alpha, \beta)$ on a non-commutative symplectic space.

Having used an abstract mathematical structure to establish the uniqueness of the Schr\"{o}dinger representation, physicists were handed a convenient mathematical tool using differential operators and wave functions, with which they were familiar and which enabled them to make calculations more easily.  Naturally the representational structure became established and accepted, independently of how it arose.

It was only later that Moyal~\cite{jm49} identified this symplectic space with a generalised phase space by identifying $\alpha=x, \beta=p$ and interpreting the algebra\footnote{Notice the Moyal algebra is isomorphic to the quantum formalism and not, as sometimes interpreted, an approximation to the quantum formalism. It should strictly be called the von Neumann-Moyal algebra.}  as providing a description of a generalised non-commutative statistics.  Unfortunately Moyal's approach was misunderstood and it became known as a semi-classical approach even though the mathematical structure it uses is exactly the quantum formalism.   

One of the key results of the von Neumann-Moyal algebra is that the quantum expectation value of an operator $\hat A$ is given by the relation
\begin{eqnarray}
\langle\psi|A|\psi\rangle=\int \int a(x,p)F_\psi(x,p)dxdp	\label{eq:expo}
\end{eqnarray}
where $F_\psi(x,p)$ is the Fourier transform of the Wigner function.  The algebra provides an explanation of the exact mathematical origins of the Wigner function.   The $a(x,p)$ are functions in $\mathbb R^{2N}$ and are subject to the non-commutative twisted or Moyal product.  These functions completely characterise the quantum state of a {\em single particle} and {\em a priori} are not specifically a feature of many-body statistics as is often assumed.

The form of equation (\ref{eq:expo}) suggests that we can interpret $F_\psi(x,p)$ as a probability density, albeit in a non-commutative space. This is indeed the way Moyal suggested it should be interpreted.  However it turns out that it is not a positive definite quantity, taking negative values in the regions where interference takes place.  This adds to the belief that the Wigner function approach is some quasi-classical approximation to quantum theory.  But the above results show this conclusion in not correct.  The von Neumann-Moyal algebra is central to the quantum formalism.  Indeed Bohm and Hiley~\cite{dbbh81} and Hiley~\cite{bh11}
have shown  how a two-point density matrix of configuration space  can be transformed into the density function $F_\psi(x,p)$, again showing that this  approach is an exact alternative formulation of the operator approach.  The Hilbert space formalism is merely a representation of this deeper structure.

\subsection{Non-locality and the Wigner Function}

Realising the significance of the Wigner function, John ~\cite{jb86}  suggested that the two-body Wigner function might provide another way of exploring the original EPR  $x-p$ correlations directly rather than only looking at spin correlations.
His analysis was based on the assumption that all quantum interference effects will produce negative values for the Wigner function.  The question would then be simply to examine the Wigner function constructed from the original  EPR wave function
\begin{eqnarray*}
\psi_\delta=\delta\left((x_1+a/2)-(x_2-a/2)\right).
\end{eqnarray*}
He found that the Wigner function was nowhere negative, implying under his assumption that there is no non-locality problem in the phase space approach. 

 However earlier Bohm and Hiley~\cite{dbbh75} 
had shown that if, for mathematical convenience, we replace the delta function by a real function $\psi$ that is very sharply peaked at $x_1-x_2=a$, the  quantum potential  takes the form
\begin{eqnarray*}
Q_\psi=-\frac{\hbar^2}{2m}\frac{\nabla^2f(x_1-x_2)}{f(x_1-x_2)}.
\end{eqnarray*}
This expression shows there is a non-local connection between the two particles so they are coupled and not free. Clearly something is wrong as both the Moyal and Bohm approaches give the same quantum expectation values for all operators.

My reasoning was that if two seemingly very different formalisms lead to the same expectation values, then there must be a relation between the two approaches.  I returned to a closer examination of the Moyal paper~\cite{jm49} and discovered that the key formulae of the Bohm interpretation already appeared in the appendix.  

 The Bohm momentum, $p_\psi=\nabla S$, is identical to a momentum defined by Moyal as a conditional expectation value 
\begin{eqnarray}
p_\psi=\int pF_\psi(x,p)dp.
\end{eqnarray}
While the transport of this momentum produced the one-body equivalent to equation (\ref{eq:2body})
\begin{eqnarray}
\partial_t S+\frac{(\nabla S)^2}{2m} + Q+V=0		\label{eq:1body}
\end{eqnarray}
where\begin{eqnarray}
Q(x, t)=-\frac{\hbar^2}{2m}\frac{\nabla^2R(x,t)}{R(x,t)}. \label{eq:1QPE}
\end{eqnarray}
These results show that, once again, there is a deep connection between the mathematics used in the Bohm approach and  the approach based on the von Neumann-Moyal algebra. 

In order to see this connection in a wider context,  we need to be aware of some important aspects of the von Neumann-Moyal algebra that have been discussed by Crumeyrolle~\cite{ac90}.  He 
has shown that the von Neumann-Moyal algebra is actually a symplectic Clifford algebra, the algebraic analogue of the more familiar orthogonal Clifford. In other words it is the geometric algebra of a symplectic space containing a Clifford group that describes the double cover of the symplectic group, namely,  the metaplectic group and its non-linear extension.  Thus the metaplectic group and its generalisation plays an analogous role to the spin groups, namely it is the covering group of the symplectic group.  It is this covering group that describes the global features of the symplectic geometry.  This once again suggests that quantum non-locality is a global feature of the dynamics that quantum processes exploit.

What is even more important from our point of view is that the Schr\"{o}dinger equation appears as the lift onto the covering space of the classical Hamiltonian flow in the symplectic space~\cite{vgss84, mdgbh11} and it is 
in the covering space that the global properties of the geometry appear.  This suggests  that the non-local properties of quantum phenomena emerge from the global properties of the covering groups in general, opening up another avenue for exploring quantum non-locality.

\section{Conclusion}

 I am forever grateful to John Bell for his tireless energy in drawing the world's attention to quantum non-locality, this truly radical feature of Nature.  I saw him forcefully defending his views against the attacks of the more conservative elements of the physics community.  I admired his courage. I also gained energy for my own investigations from his fierce defence of the right of physicists to investigate further the full implications of different approaches to quantum phenomena, approaches that had been wrongly criticised by the founding fathers of quantum mechanics.
 Whenever I met him on conferences at various venues in Europe, he would always be encouraging, defending the right to critically explore quantum phenomena in new ways.  I certainly needed that encouragement.  John, you left us too soon, but thank you for your support.


\end{document}